\newif\ifARXIV
\ARXIVtrue

\ifARXIV
\documentclass[sigconf,nonacm]{acmart}
\else
\documentclass[sigconf,review]{acmart}
\fi
\AtBeginDocument{%
  }

\setcopyright{acmlicensed}
\copyrightyear{2018}
\acmYear{2018}
\acmDOI{XXXXXXX.XXXXXXX}
\acmConference[Conference acronym 'XX]{Make sure to enter the correct
  conference title from your rights confirmation email}{June 03--05,
  2018}{Woodstock, NY}
\acmISBN{978-1-4503-XXXX-X/2018/06}

\usepackage{graphicx}
\usepackage{textcomp}
\usepackage{listings}
\usepackage{xcolor}
\usepackage{comment}
\usepackage{array}
\usepackage[skins,breakable,listings]{tcolorbox}

\usepackage[frozencache,cachedir=.]{minted}

\lstdefinestyle{promptstyle}{
  basicstyle=\small\ttfamily,
  breaklines=true,
  columns=flexible,
  keepspaces=true,
  showstringspaces=false,
  tabsize=2
}

\newtcblisting{promptbox}[2][]{%
  enhanced,
  breakable,
  listing only,
  listing options={style=promptstyle},
  title={#2},
  colback=gray!3,
  colframe=black!20,
  boxrule=0.4pt,
  arc=1mm,
  left=1mm,right=1mm,top=1mm,bottom=1mm,
  fonttitle=\bfseries\color{black}, 
  fontsize=\small,
  coltitle=black, 
  #1
}

\newcommand{\myprompt}[2]{%
  \begin{tcolorbox}[
    enhanced,
    colback=gray!10,
    colframe=black!20,
    boxrule=0.4pt,
    arc=1mm,
    left=2mm,right=2mm,top=1mm,bottom=1mm,
    fonttitle=\bfseries,
    coltitle=black,
    title={#1},
    before skip=3pt plus 2pt minus 1pt,
    after skip=3pt plus 2pt minus 1pt
  ]
  \small #2
  \end{tcolorbox}%
}

\newif\ifDEBUG
\DEBUGtrue
\DEBUGfalse

\newif\ifTABLEFONTHACK
\TABLEFONTHACKtrue
\TABLEFONTHACKfalse

\newif\ifDESIGN
\DESIGNfalse

\usepackage{algpseudocode}
\usepackage[normalem]{ulem} 
\usepackage{xspace}
\usepackage{multirow} 
\usepackage{balance} 
\usepackage{fancyvrb} 
\usepackage{caption}
\usepackage{booktabs} 

\usepackage{enumitem}
\setlist[itemize]{leftmargin=*,noitemsep,topsep=0pt}
\setlist[enumerate]{leftmargin=*}

\usepackage{etoolbox}
\makeatletter
\patchcmd{\@makecaption}
	{\scshape}
	{}
	{}
	{}
\makeatletter
\patchcmd{\@makecaption}
	{\\}
	{.\ }
	{}
	{}
\makeatother

\newcommand{\ie}{\textit{i.e.,}\xspace}
\newcommand{\eg}{\textit{e.g.,}\xspace}
\newcommand{\etal}{\textit{et al.}\xspace}

\usepackage{mdframed}
\mdfsetup{skipabove=0.5\topskip,skipbelow=0.5\topskip,align=center}

\usepackage{amsthm}
\newtheorem{thm}{Theorem}

\DeclareMathSymbol{\mlq}{\mathord}{operators}{``}
\DeclareMathSymbol{\mrq}{\mathord}{operators}{`'}

\newif\ifSAVESPACE
\SAVESPACEfalse

\ifSAVESPACE

\else

\fi

\usepackage{todonotes}
\usepackage{soul}
\usepackage{fontawesome}
\usepackage{ragged2e}

\ifDEBUG
    \newcommand{\AH}[1]{\todo[color=cyan,inline]{AH:#1}}
    \newcommand{\AM}[1]{\todo[color=red,inline]{Machiry:#1}}
    \newcommand{\JD}[1]{\todo[color=yellow,inline]{JD:#1}}
    \newcommand{\TL}[1]{\todo[color=green,inline]{SA:#1}}
    \newcommand{\PA}[1]{\todo[color=orange,inline]{PA:#1}}
    
    \newcommand{\KR}[1]{\todo[color=yellow,inline]{Kyle:#1}}
    \newcommand{\LS}[1]{\todo[color=green,inline]{LS:#1}}
    \newcommand{\HP}[1]{\todo[color=green,inline]{HP:#1}}
    \newcommand{\PP}[1]{\todo[color=lime,inline]{PP: #1}}

\else
    \newcommand{\AH}[1]{}
    \newcommand{\AM}[1]{}
    \newcommand{\JD}[1]{}
    \newcommand{\TL}[1]{}
    \newcommand{\PA}[1]{}
    \newcommand{\KR}[1]{}
    \newcommand{\LS}[1]{}
    \newcommand{\HP}[1]{}
    \newcommand{\PP}[1]{}

\fi

\usepackage{hyperref}

\usepackage{cleveref}
\crefformat{section}{\S#2#1#3}
\crefname{figure}{Figure}{Figures}
\crefname{table}{Table}{Tables}
\crefname{theorem}{Theorem}{Theorems}
\crefname{thm}{Theorem}{Theorems}
\crefname{lemma}{Lemma}{Lemmata}
\crefname{equation}{Eqt.}{Eqts.}
\crefformat{Grammar}{Grammar #1}
\crefname{appendix}{Appendix}{Appendices}
\crefname{listing}{Listing}{Listings}

\newcommand{\myparagraph}[1]{\paragraph{#1}}
\renewcommand{\myparagraph}[1]{\vspace{0.25em} \noindent \hspace{0.1cm}\underline{\textit{#1:}}}

\usepackage{url}

\usepackage{tcolorbox}

\makeatletter
\newcommand{\linebreakand}{%
  \end{@IEEEauthorhalign}
  \hfill\mbox{}\par
  \mbox{}\hfill\begin{@IEEEauthorhalign}
}
\makeatother

\usepackage{pifont}

 \usepackage{pifont}

\begin{document}

\newcolumntype{P}[1]{>{\centering\arraybackslash}p{#1}}
\newcommand{\OF}{OSS-Fuzz\xspace}
\newcommand{\OFG}{OSS-Fuzz-Gen\xspace}

\title{On Mitigating False Crashes in \OFG: An Experience Report}
\title{LLM-Driven False Crash Mitigation in Fuzz Driver Generation}
\title{Program Analysis Agents to Mitigate False Positive Crashes during Function-level Fuzzing}
\title{FalseCrashReducer: Agentic AI-Driven Strategies to Mitigate False Positive Crashes in \OFG}
\title{FalseCrashReducer: Mitigating False Positive Crashes in \OFG Using Agentic AI}

\author{Paschal C. Amusuo}
\affiliation{%
  \institution{Purdue University}
  \country{USA}
}
\email{pamusuo@purdue.edu}

\author{Dongge Liu}
\affiliation{%
  \institution{Google LLC}
  \country{Australia}
}
\email{donggeliu@google.com}

\author{Ricardo Calvo}
\affiliation{%
  \institution{Purdue University}
  \country{USA}
}
\email{rcalvome@purdue.edu}

\author{Jonathan Metzman}
\affiliation{%
  \institution{Google LLC}
  \country{USA}
}
\email{metzman@google.com}

\author{Oliver Chang}
\affiliation{%
  \institution{Google LLC}
  \country{Australia}
}
\email{ochang@google.com}

\author{James C. Davis}
\affiliation{%
  \institution{Purdue University}
  \country{USA}
}
\email{davisjam@purdue.edu}

\begin{abstract}
Fuzz testing has become a cornerstone technique for identifying software bugs and security vulnerabilities, with broad adoption in both industry and open-source communities. 
Directly fuzzing a function requires fuzz drivers, which translate random fuzzer inputs into valid arguments for the target function. 
Given the cost and expertise required to manually develop fuzz drivers, methods exist that leverage program analysis and Large Language Models to automatically generate these drivers.
However, the generated fuzz drivers frequently lead to false positive crashes, especially in functions highly structured input and complex state requirements. 
This problem is especially crucial in industry-scale fuzz driver generation efforts like \OFG, as reporting false positive crashes to maintainers impede trust in both the system and the team.

This paper presents two AI-driven strategies to reduce false positives in \OFG, a multi-agent system for automated fuzz driver generation. 
First, \emph{constraint-based fuzz driver generation} proactively enforces constraints on a function’s inputs and state to guide driver creation. 
Second, \emph{context-based crash validation} reactively analyzes function callers to determine whether reported crashes are feasible from program entry points. 
Using 1,500 benchmark functions from \OF, we show that these strategies reduce spurious crashes by up to 8\%, cut reported crashes by more than half, and demonstrate that frontier LLMs can serve as reliable program analysis agents. 
Our results highlight the promise and challenges of integrating AI into large-scale fuzzing pipelines.
\end{abstract}



\maketitle


\section{Introduction}

Fuzzing~\cite{miller_empirical_1990, manes_art_2021}, or fuzz testing, is a key software engineering technique for uncovering bugs and security vulnerabilities. 
It is widely used in both commercial~\cite{smith_browser_2021, marinescu_autonomous_2021, noauthor_gentle_2019} and open-source projects~\cite{serebryany_oss-fuzz_2017}. 
To target specific application functions or libraries, engineers create fuzz drivers that convert random fuzzer inputs into the structured arguments expected by the target functions. 
However, manually writing fuzz drivers for a project requires project-level expertise, is labor-intensive and error-prone, leading to limited fuzzing coverage even in continuously fuzzed projects~\cite{noauthor_fuzzing_nodate}.

Several research efforts have explored automatically generating fuzz drivers for project functions. 
These approaches either rely on program analysis~\cite{zhang_apicraft_2021, babic_fudge_2019, ispoglou_fuzzgen_2020, sherman_no_2025} or LLM-based techniques~\cite{Liu_OSS-Fuzz-Gen_Automated_Fuzz_2024, xu_ckgfuzzer_2025, lyu_prompt_2024}. 
Google’s \OFG~\cite{Liu_OSS-Fuzz-Gen_Automated_Fuzz_2024} is an ongoing project that applies LLMs to generate drivers at scale for critical open-source software. 
A persistent challenge, however, is that automatically generated drivers can produce invalid inputs and result in false positive crashes that do not correspond to real bugs. 
Muralee~\etal~\cite{muralee_reactive_2025} note this problem is intrinsic to bottom-up testing, where non-entry-point functions are directly fuzzed, as these functions typically expect well-structured and validated inputs. 
Existing solutions for detecting these false positive crashes either rely on program analysis techniques with high engineering complexity~\cite{muralee_reactive_2025, liu_afgen_2024} or imprecise LLM-based strategies~\cite{xu_ckgfuzzer_2025}, limiting applicability in large-scale settings like \OFG that handles thousands of projects. 

In this paper, we propose and evaluate FalseCrashReducer, two LLM-driven strategies to mitigate false positive crashes in \OFG. 
The first is a proactive crash reduction strategy, \emph{constraint-based fuzz driver generation}, which derives and applies constraints on a function's inputs and state to guide fuzz driver creation. 
The second is a reactive crash reduction strategy, \emph{context-based crash validation}, that analyzes a function's callers to determine whether a crash can be triggered when the project is executed from its entry point. 
To implement these strategies, we design two LLM-based agents, the \emph{function analyzer agent} and the \emph{crash validation agent}, and integrate them into \OFG.

We evaluate the impact and cost of the proposed strategies in \OFG using 1,555 benchmark functions from the \OF framework. 
Our findings show that 
 (1) constraint-based fuzz driver generation reduces the number of crashes by 2--8\%, with 24.2\% more fuzz drivers respecting the target function's constraints; 
 (2) context-based crash validation reduces the number of reported crashes by 57.3 -- 61.3\%, significantly lowering the debugging burden for software engineers; and 
 (3) generating fuzz drivers with \OFG{} costs less than a dollar, with tool usage contributing the highest proportion of costs.

In summary, our contributions are:
\begin{itemize}
    \item The first description of the design and architecture of \OFG, Google's multi-agent system for creating fuzz drivers.
    \item Design and evaluation of two novel agent-driven strategies to proactively reduce and reactively filter false positive crashes.
\end{itemize}

\vspace{0.1cm}
\noindent
\uline{Significance:}
Automatic fuzz driver generation removes the bottleneck of manual driver creation, expands coverage, and improves vulnerability discovery on critical projects.
However, state-of-the-art approaches like \OFG are hindered by false positives that increase debugging effort and undermine credibility.
We identified, designed, and evaluated two complementary approaches to reduce these false positives.
Our two strategies directly enhance the usability of \OFG, which benefits critical open-source projects.
However, they are not tightly coupled to \OFG, and can be used in other automated testing systems.
Based on our experience, we identify a range of future works to advance automated software testing.
All results are open-source to facilitate broad review and adoption.


\section{Background and Related Work}

\subsection{Function-Level Fuzzing and Fuzz Drivers}
\label{subsec:bg-fuzz-drivers}

\begin{listing}
  \centering
  \caption{Implementation of \texttt{crxDecodePlane} in \texttt{libraw} library.
  It expects input pointer \texttt{p} to reference a well-formed \texttt{CrxImage} object, derefencing without validity checks.
  }
  \label{listing:crxDecodePlane_source}
\begin{minted}[
    fontsize=\scriptsize,
    linenos,
    gobble=2, % Remove unnecessary indentation -- the line numbers make this clear enough
    xleftmargin=0.5cm, % Otherwise we start in the left margin...
    escapeinside=||, % Allows you to use LaTeX commands inside code
    style=colorful,   % You can change this to another style like 'monokai', 'borland', etc.
    breaklines        % Break long lines
]{c}

int LibRaw::crxDecodePlane(void *p, uint32_t planeNumber) {
  CrxImage *img = (CrxImage *)p;
  for (int tRow = 0; tRow < img->tileRows; tRow++) {
    for (int tCol = 0; tCol < img->tileCols; tCol++) {
      CrxTile *tile = img->tiles + tRow * img->tileCols + tCol;
      CrxPlaneComp *planeComp = tile->comps + planeNumber;
      uint64_t tileMdatOffset = tile->dataOffset + tile->mdatQPDataSize + tile->mdatExtraSize + planeComp->dataOffset;
        ...
    }
  }
}
\end{minted}
\end{listing}

\begin{listing}
  \centering
  \caption{
  Fuzz driver for the \texttt{crxDecodePlane} function (\cref{listing:crxDecodePlane_source}).
  The fuzz driver violates the preconditions of the target function, leading to false positive crashes.
  }
  \label{listing:fuzzing-driver}
\begin{minted}[
    fontsize=\scriptsize,
    linenos,
    gobble=2, % Remove unnecessary indentation -- the line numbers make this clear enough
    xleftmargin=0.5cm, % Otherwise we start in the left margin...
    escapeinside=||, % Allows you to use LaTeX commands inside code
    style=colorful,   % You can change this to another style like 'monokai', 'borland', etc.
    breaklines        % Break long lines
]{c}

extern "C" int LLVMFuzzerTestOneInput(const uint8_t *data, size_t size) {
  FuzzedDataProvider provider(data, size);

  uint32_t planeNumber = provider.ConsumeIntegralInRange<uint32_t>(0, 3);
  size_t crx_image_size = provider.ConsumeIntegralInRange<size_t>(0, 2048);
  std::vector<uint8_t> crx_image_buf = provider.ConsumeBytes<uint8_t>(crx_image_size);

  // Call the public wrapper which in turn calls the protected target function.
  RawProcessor.crxDecodePlane(crx_image_buf.data(), planeNumber);
  return 0;
}
\end{minted}
\end{listing}

Fuzzing~\cite{manes_art_2021} is a software testing technique that exercises programs with randomly generated or mutated inputs to discover bugs or security vulnerabilities. 
According to Muralee~\etal~\cite{muralee_reactive_2025}, fuzzing can either be \textit{top-down} or \textit{bottom-up}.
Top-down fuzzing~\cite{fioraldi_afl_2020, stephens_driller_2016} fuzzes a program from its public entry point, feeding inputs via expected channels such as command-line arguments, input files, or standard input. 
Although this approach has high validity --- crashes detected during execution usually indicate real software bugs --- it struggles with low coverage on complex or deeply nested paths.

OSS-fuzz-gen takes the alternative approach, \textbf{\textit{bottom-up fuzzing}} \cite{liu_afgen_2024, muralee_reactive_2025}.
This strategy targets individual functions, which may not be public entry points. 
It is used to 
exercise deep or rare functions, improving coverage and fuzzing efficiency.

To perform bottom-up fuzzing, developers create \emph{fuzz drivers}~\cite{babic_fudge_2019,ispoglou_fuzzgen_2020}. 
These are programs that invoke the target function(s) using fuzzer-generated inputs. 
The driver must
  ensure that the function is called in an environment resembling normal execution,
  by
  setting up the required context (\eg global and module-specific state)
  and
  constructing valid input structures.
Additional initialization and cleanup in the driver reduce errors resulting from incorrect resource management.
For example,~\cref{listing:fuzzing-driver} illustrates a fuzzing driver for the function \texttt{crxDecodePlane}. 
The driver translates raw fuzzer inputs from \texttt{provider} into arguments for the target function and handles necessary setup and teardown operations.

Because fuzz drivers can directly target intermediate functions in the program, they may trigger crashes that would never arise in real executions. 
These \textit{false positive crashes} are the central challenge in function-level fuzzing~\cite{muralee_reactive_2025}. 
To illustrate, consider again the fuzz driver shown in~\cref{listing:fuzzing-driver}.
This driver invokes the target using a buffer of arbitrary size. 
In the target (\cref{listing:crxDecodePlane_source}), that buffer, \texttt{*p}, is cast to a \texttt{CrxImage} object and accessed. 
If the buffer is too small, a crash occurs due to invalid memory access. 
If all real callers of \texttt{crxDecodePlane} are well-formed, then this crash is a false positive. 
These false positive crashes increase debugging burden, obscure genuine bugs, and erode project maintainers' trust.


\subsection{Automatically Generating Fuzz Drivers}
\label{subsec:bg-generating-fuzz-drivers}

It is costly to manually develop a fuzz driver for every function of interest. 
Researchers have therefore proposed methods for automatically generating fuzz drivers. 

\textit{Program-Analytic approaches}~\cite{babic_fudge_2019, ispoglou_fuzzgen_2020, sherman_no_2025, liu_afgen_2024, muralee_reactive_2025} create fuzz drivers systematically and rely on program analysis to infer the inputs and context needed to invoke target functions. 
They leverage different techniques including program slicing~\cite{babic_fudge_2019}, model-based~\cite{ispoglou_fuzzgen_2020} and type-based~\cite{muralee_reactive_2025} construction methods to produce fuzz drivers that compile and target the required functions. 
However, while correct by construction and leading to lower false positive crashes, they are typically complex by design and require substantial engineering effort to implement correctly.

Recently, \textit{AI-based approaches}~\cite{xu_ckgfuzzer_2025, lyu_prompt_2024, zhang_how_2024, Liu_OSS-Fuzz-Gen_Automated_Fuzz_2024} have emerged, leveraging large language models (LLMs) to generate fuzz drivers. 
These models can create more diverse drivers, combining functions in novel ways beyond existing consumer patterns. 
However, their reliance on AI also increases the likelihood of fuzz driver errors, producing false positive crashes during fuzzing and imposing additional debugging overhead on engineers.

These prior works have also integrated different ways to filter out false positive crashes. 
This includes using program analysis techniques like symbolic execution~\cite{muralee_reactive_2025} and static constraint analysis~\cite{liu_afgen_2024} to validate validate the feasibility of crashes, using LLMs to detect invalid crashes based on crash locations and patterns~\cite{xu_ckgfuzzer_2025}, and the use of heuristics~\cite{lyu_prompt_2024} to identify potentially false positive crashes. 
Yet, program-analysis-based methods are often too complex to scale across diverse software projects, and current AI or heuristics-based methods do not utilize whole-program context, reducing their precision.

This gap highlights the need for new strategies that are easily applicable across diverse software projects, and capable of incorporating whole-program context to more effectively reduce false positive crashes.

\section{Context: \OF and \OFG}

This section describes the industry context of our work.
As there is no academic material on \OFG, we describe the system in enough detail that the reader understand our contribution to false positive mitigation within it and the open problems.
However, we defer a detailed evaluation of its design choices to a future paper.


\subsection{\OF: Fuzzing Framework for OSS}
\label{subsec:context-oss-fuzz}

\OF~\cite{noauthor_announcing_nodate} is Google’s continuous fuzzing service designed to uncover security vulnerabilities and improve the reliability of critical open-source software (OSS). 
It currently supports more than 1,300 projects selected for their widespread use or importance to global IT infrastructure~\cite{noauthor_accepting_nodate}. 
As of May 2025, \OF has helped identify and fix over 13,000 security vulnerabilities and 50,000 bugs across 1,000 projects. 
It has also been a subject of many academic studies, including studies on fuzzing performance~\cite{gorz_empirical_2025, gao_beyond_2023, nourry_my_2025}, bugs~\cite{ding_empirical_2021, keller_what_2023}, and automation~\cite{zhang_fixing_2024, zhang_how_2024}
Complementing its core service, \OF provides the Open Source Fuzzing Introspection platform~\cite{noauthor_fuzzing_nodate}, which leverages Fuzz Introspector~\cite{noauthor_ossffuzz-introspector_2025} to analyze project fuzzing performance and make results available via a public website and API~\cite{noauthor_fuzz_nodate}.

\OF projects rely on fuzz drivers that exercise specific functions or subsystems within a codebase. 
However, writing high-quality drivers that achieve broad coverage is time-intensive and requires deep domain expertise. 
Consequently, many projects still exhibit significant coverage gaps. 
Gao~\etal~\cite{gao_beyond_2023} showed that most of these gaps stem from limitations in existing drivers, highlighting the need for more effective ones. 
To address this, the \OF team has begun exploring automated techniques for fuzz driver generation to expand coverage and improve bug discovery~\cite{liu_ai-powered_nodate}.



\subsection{\OFG: Fuzz Driver Generation}
\label{subsec:context-oss-fuzz-gen}

\OFG~\cite{Liu_OSS-Fuzz-Gen_Automated_Fuzz_2024} is a multi-agent system developed by the \OF team to automate fuzz driver generation and evaluation for open-source projects. 
Though still under development, it has already uncovered 30 previously unknown bugs and vulnerabilities~\cite{Liu_OSS-Fuzz-Gen_Automated_Fuzz_2024} and delivered major coverage improvements in projects, including a 98.42\% coverage gain in \texttt{phmap}~\cite{Liu_OSS-Fuzz-Gen_Automated_Fuzz_2024}.

\myparagraph{Design and Architecture}  
\OFG employs an LLM-based agentic approach to generate fuzz drivers for functions with little or no coverage. 
This bottom-up focus (\cref{subsec:bg-fuzz-drivers}) targets functions deep in a call-graph that typically require structured inputs or program states from higher-level code. 
Because creating drivers at this level requires reasoning about code semantics, dependencies, and input constraints, LLM-based agents are well suited for the task. 
Their reasoning ability also enables \OFG to generalize across diverse \OF projects with minimal manual effort.  

\begin{figure}[h]
    \centering
    \includegraphics[width=\linewidth]{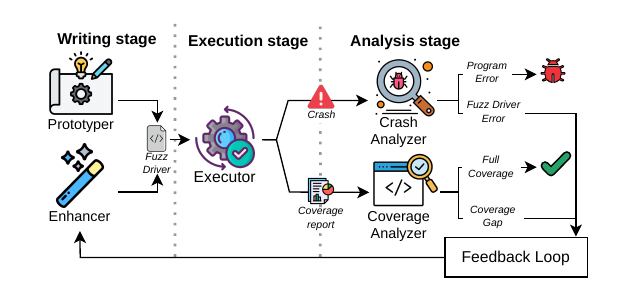}
    \caption{
    OSS-Fuzz-Gen design showing its agents.
    A bottom-up approach is taken, targeting functions with low coverage.
    This design exhibits a high false positive rate.
    }
    \label{fig:oss-fuzz-gen-design}
\end{figure}

As shown in \cref{fig:oss-fuzz-gen-design}, \OFG organizes multiple agents into three stages executed in a pipeline:  
(1) \textit{writing stage}, where drivers are generated for target functions;  
(2) \textit{execution stage}, where drivers are fuzzed; and  
(3) \textit{analysis stage}, where execution results are analyzed and used to guide fuzz driver refinements.  
Agents are equipped with task-specific instructions and tools (\cref{tab:oss-fuzz-gen-agents}), and the system integrates a feedback cycle, stopping after (1) a true positive bug; (2) a maximum number of cycles; or (3) a coverage plateau. 

This coverage-guided bottom-up strategy contrasts and complements prior fuzz driver generation approaches~\cite{babic_fudge_2019, ispoglou_fuzzgen_2020, xu_ckgfuzzer_2025, lyu_prompt_2024} that target public library APIs and explores random API combinations. 
Compared to these approaches, \OFG is well suited for the existing \OF projects that already contain fuzz drivers for their public APIs but still suffer from low coverage.
However, as shown by \cite{muralee_reactive_2025}, this bottom-up approach introduces higher risk of false positive crashes caused by bypassing normal entry points which could have validated malformed input.

\myparagraph{Implementation}  
\OFG is designed for scale, intended to support thousands of projects (\OF currently fuzzes 1311 projects~\cite{noauthor_fuzzing_nodate}). 
It runs on Google’s distributed cloud infrastructure, with each agent isolated in a container and managed by a central orchestrator. 
This design enables parallel execution and cross-project fault tolerance. 
The project's implementation is publicly available~\cite{Liu_OSS-Fuzz-Gen_Automated_Fuzz_2024} and comprises $\sim$24,000 lines of Python.

Additionally, agents communicate through a pipe-and-filter architecture~\cite{claytonsiemens77_pipes_nodate}, passing outputs directly to the next stage. 
This mechanism sufficed when the output of one agent was only consumed by the next pipelined agent, although we needed to change it when we introduced the function analyzer agent (\cref{subsubsec:ofg-improvements}).  

\begin{table}
    \centering
    \small
    \caption{
    Existing \OFG agents, Description and tools.
    }
    \begin{tabular}{p{1.2cm}p{4.5cm}p{1.6cm}}
        \toprule
        \textbf{Agents} & \textbf{Description} & \textbf{Tools} \\
        \midrule
        Prototyper & Creates the initial fuzz driver. & Compiler, Code search \\
        Enhancer & Refines fuzz driver w/ analysis feedback. & Compiler, Code search \\
        Coverage Analyzer & Analyzes coverage reports and makes suggestions to improve coverage. & Code search \\
        Crash Analyzer & Triages crashes and classifies them as program or fuzz driver error. & Code search, Debugger \\
        \bottomrule
    \end{tabular}
    \label{tab:oss-fuzz-gen-agents}
\end{table}

\subsection{Problem: False Positive Crashes}

\OFG's bottom-up approach makes it prone to false positive crashes. 
In our evaluation (\cref{tab:rq1-crashes}), 1555 benchmarks produced 4835 crashes (averaging 3.1 per benchmark), 70\% of which were marked false positives. 
Addressing this issue is critical: reporting false positives to maintainers of widely used, security-critical \OF projects undermines the credibility of both \OFG and the \OF team, and delays the resolution of defects.  

To mitigate this problem, the \OFG team has developed crash triage and classification tools. 
An early \textit{semantic analyzer}, based on recommendations from Zhang~\etal~\cite{zhang_how_2024}, combined generative AI with heuristics to validate crashes, and more recently a \textit{crash analyzer agent} was introduced that applies debugging tools to inspect program state, identify root causes, and classify crashes as either \textit{``Program Errors''} or \textit{``Fuzz Driver Errors''}. 
While useful for post-crash analysis, these tools do not proactively prevent false positives nor incorporate broader program context, limiting precision.  

In this work, we investigate new context-based strategies to more effectively reduce false positive crashes in \OFG.  



\section{Designs to Mitigate False Positive Crashes}

This section presents our agent-based designs to reduce false positive crashes in \OFG.

\subsection{Problem Statement and Goal}

\ifARXIV
\myparagraph{Problem Statement}
Fuzz drivers targeting intermediate functions in a program often produce false positive crashes.  
These arise when the fuzz drivers generate inputs that are not feasible in normal execution. 
They increase debugging overhead and reducing the usability and credibility of fuzz driver generation systems.

\myparagraph{Goal}
\fi
This project aims to design, implement, and evaluate strategies to reduce or filter false positive crashes during fuzz driver generation, which can be integrated into \OFG's agent pipeline, and can scale to the diverse open-source projects on \OF.


\subsection{Design Overview}

We adopt two complementary strategies to reduce false positive crashes:  
proactive crash reduction and reactive crash validation.

\begin{itemize}
    \item \textit{Constraint-based Fuzz Driver Generation:}  
    This proactive strategy derives constraints on how a target function should be used and applies them during fuzz driver generation in \OFG.  
    By enforcing correct function usage, it reduces invalid fuzz drivers that cause false positives.

    \item \textit{Context-based Crash Validation:}  
    This reactive strategy validates crashes flagged as program errors by checking if they can be triggered from public entry points during normal execution.
\end{itemize}

Both strategies are necessary. 
Proactive reduction reduces invalid drivers and lowers validation costs, but attempting to eliminate all false positives with overly strict constraints risks missing real bugs.  
Reactive validation is necessary to balance precision with bug-finding effectiveness.

\begin{figure}[h]
    \centering
    \includegraphics[width=1\linewidth]{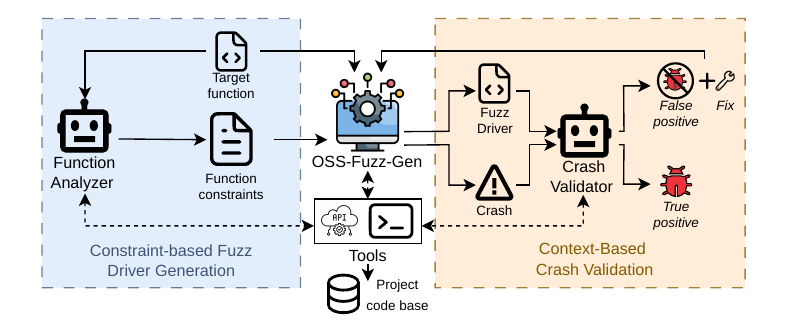}
    \caption{
    Agent-driven strategies to mitigate false positive crashes in \OFG (cf. \cref{fig:oss-fuzz-gen-design}).
    Semantic constraints developed by function analyzer improve fuzz driver quality and prevent false crashes. 
    The crash validator analyzes project context to determine crash's feasibility and filter false positives. 
    Agents use tools to access the project’s codebase.
    }
    \label{fig:approach-design}
\end{figure}

We design LLM-based agents with access to the project’s source code to derive function requirements (proactive) and validate crash feasibility (reactive) and integrated them into \OFG’s distributed workflow (\cref{fig:approach-design}).  
Following the agent framework in~\cite{noauthor_llm_nodate}, we describe each agent’s input, reasoning, tools, and output. We omit planning and memory modules since frontier LLMs provide them implicitly.

\subsection{Part 1: Constraint-based Driver Generation}

\subsubsection{Rationale}
False positive crashes typically occur when fuzz drivers incorrectly initialize the state or input of a function before calling it.
An example is shown in \cref{listing:fuzzing-driver} where random input bytes is used to call a function that expects a valid CrxImage object.
Programmatic fuzz driver generation methods develop constrained fuzz drivers by default, as the fuzz drivers mimick existing real-world code.
However, existing AI-based methods remain unconstrained, relying on the LLM to determine how functions in the fuzz driver should be called.
Hence, to balance LLM flexibility with correctness, we introduce the use of function constraints, representing precise instructions for calling the target function correctly, to guide the LLM when generating drivers. 

\subsubsection{Function Constraints}
These are instructions that define how to correctly setup a function's state and input arguments before the function is called.
We identify four categories of constraints, representing the conditions we observed that frequently led to incorrect fuzz drivers.

\begin{itemize}
    \item \textit{Input construction methods:} Instructions for creating input variables. This includes what functions to use to create these variables or if they can be directly initialized with fuzz data.
    \item \textit{Variable constraints:} Bounds and preconditions on input variables. This include ranges for scalar variables or buffer sizes necessary to satisfy array indexing or assertion conditions, and null pointer and termination conditions for pointer and string variables.
    \item \textit{Input relationships:} Expected dependencies between variables used by the function, such as the connection between a pointer and its associated size field.
    \item \textit{Setup and teardown functions:} Functions that must be called before or after the target function to ensure correct initialization and cleanup of the calling state or global variables used by the target function.
\end{itemize}

Below, we show the constraints derived for the function in \cref{listing:crxDecodePlane_source}.
As shown, these constraints capture the requirements sufficient to avoid the surface level crashes that would be caused by the fuzz driver in \cref{listing:fuzzing-driver}.

\myprompt{Some constraints derived for \cref{listing:crxDecodePlane_source}}{
\begin{itemize}
\item The first argument must be a valid pointer to a 'CrxImage' structure. This is because...
\item 'crxSetupImageData' function must be called before 'crxDecodePlane'. This is because...
\item 'planeNumber' must be less than 'nPlanes' because...
\item 'CrxImage' structure should be initialized by 'crxLoadRaw'...
\end{itemize}
}

\subsubsection{Function Analyzer Agent Design}
\label{subsubsec:function-analyzer-design}
To automatically derive these requirements, we design a Function Analyzer Agent capable of analyzing a function and its usages within a broader project to derive the expected requirements. 

\myparagraph{Agent's Input}
The Function Analysis Agent is provided with details of the target function, including the containing project's name, the function's signature, and the source code of the function, together with a path to the location of the project's codebase.

\myparagraph{Agent's Tools}  
We integrate two tools that allow the agent to explore and analyze a project’s codebase.  
We provide detailed instructions and examples of valid tool usage in the agent’s prompt to guide effective interaction. 

\begin{itemize}  
    \item \textit{Code Search Tool:}
    This tool enables the LLM to search the project’s codebase using Linux shell commands.

    \item \textit{Function Search Tool:} This tool retrieves the implementation of a specific function by querying the Fuzz Introspector API service~(\cref{subsec:context-oss-fuzz}). The LLM provides the project and function name, and the tool returns the implementation if it has been indexed. This tool is more efficient for obtaining complete function definitions. 
\end{itemize}  

Prior work on agents for program analysis also supports code search but typically abstracts away shell commands or tool details~\cite{liu_agent_2025, zhang_autocoderover_2024}.
Our approach gives the LLM flexibility to issue commands within an isolated environment.  

\myparagraph{Agent's Prompt}
We adopt a problem decomposition strategy~\cite{xiao_foundations_2025, khot_decomposed_2023, prasad_adapt_2024} that breaks the task into sequential steps, guiding the LLM through the constraint generation process. 
Below is a shortened version of the prompt to highlight its core structure. 

\myprompt{Function Analyzer Agent Prompt}{
You are a security engineer about to create...
Your goal is to analyze the target function and its usages, and identify constraints...

\textit{Input}: [Input items...]

\textit{Categories of function constraints}:
    [categories...]
    
\textit{Steps to follow}:
\begin{itemize}
\item Identify function's parameters and callers.
\item Determine implicit assumptions on parameters.
\item Analyze how parameters are constructed in callers.
\item Identify common setup and teardown functions.
\item Compile results into a list of constraints.
\end{itemize}

\textit{Output}: [Output format...]

\textit{Examples}: [Examples...]

\textit{Tools provided}: [Tool list...]

[Detailed Tool Instructions...]
}


\myparagraph{Agent's Output}
The agent is prompted to produce a detailed description of the provided function and a list of requirements guiding how the function should be called correctly by the fuzz driver.

\subsection{Part 2: Context-Based Crash Validation}

\subsubsection{Rationale}
While function requirements can reduce surface-level crashes caused by invalid inputs or states, some false positive crashes arise deeper within target function's call graph or in other functions invoked by the fuzz driver.  
These crashes may look genuine in isolation but are not feasible under realistic execution paths starting from a project’s external entry points.  
To filter out such spurious reports, we introduce a context-based validation strategy that determines whether a reported crash can actually be triggered within the project’s broader execution context.  

\subsubsection{Crash Feasibility}
We define a crash as \textit{feasible} if it can be triggered from a project’s external entry points, which we identify as root-level non-test functions in the project’s call graph.  
A false positive crash, by contrast, is one whose conditions cannot be satisfied by any execution path beginning at these entry points.

To determine feasibility, the crash’s triggering conditions must be reconstructed from the stacktrace and root cause analysis, and then checked against the constraints present in real calling contexts.  
This ensures that only true positive crashes, reachable from valid entry points, are retained.

\subsubsection{Crash Validation Agent Design}
\label{subsubsec:cv-agent-design}
To realize context-based crash validation, we design a Crash Validation Agent that can analyze the project and functions associated with a crash and determining if the crash is reachable from the project's entry points.
We describe this agent using the same structure as \cref{subsubsec:function-analyzer-design}.

\myparagraph{Agent's Input}
For input, the Crash Validation Agent is provided crash details:
  stacktrace,
  crash logs,
  and
  a root cause analysis produced by \OFG's Crash Analyzer Agent (\cref{fig:oss-fuzz-gen-design}).

\myparagraph{Agent's Tools}
The Crash Validation Agent uses the same tools as the Function Analyzer Agent (\cref{subsubsec:function-analyzer-design}) to explore source code.

\myparagraph{Agent's Prompt}
The crash validation agent follows a similar prompting strategy as the function analyzer agent. 
We show a shortened version that highlight the instructions and guidance provided.

\myprompt{Crash Validation Agent Prompt}{
You are a security engineer developing...Your goal is to analyze the crash details and determine if the crash is feasible from the project's entry points.

\textit{Input}: [Input items...]

\textit{Steps to follow}:
\begin{itemize}
\item Identify the crashing function and crash location.
\item Determine the input conditions that caused the the crash.
\item Identify how input arguments are created at call sites.
\item Analyze whether constraints on input arguments could have prevented the crash.
\item Provide your conclusion and code evidence for claims.
\end{itemize}

\textit{Output}: [Output format...]

\textit{Tools provided}: [Tool list...]

[Detailed Tool Instructions...]
}

\myparagraph{Agent's Output}
The agent produces a structured report containing its conclusion and analysis about the crash's feasibility, evidence from the source code, and recommendations for fuzz driver modifications for crashes identified as false positives.

\subsection{Implementation}

We implemented the Function Analysis and Crash Validation agents using Google’s Agent Development Kit (ADK). 
Their unique implementations required 254 lines of Python code.  
Because the agents were built on \OFG, they reused much of its utility code, including components for prompt preparation, LLM interaction, error handling, and tool support.
The agents’ initial prompts totaled 208 lines, excluding variable elements (\eg crash stacktraces for the Crash Validation agent) and additional reprompting prompts for error handling.  

Overall, \OFG comprises 25k lines of Python code, covering its core functionality, supporting tools (\eg report generation, agent debugging, etc.), and experimental modules. 
Within \OFG, the four existing agents (\cref{tab:oss-fuzz-gen-agents}) together account for 712 lines of code, plus 345 lines of prompt definitions.
Thus the size of the new false positive-related agents is comparable to the existing ones.


\subsubsection{Integrating Agents}
We integrate the Function Analyzer and Crash Validation agents to \OFG. 

\myparagraph{Integrating Function Analyzer Agent} 
We integrate the Function Analyzer at the start of \OFG’s pipeline.
Function constraints produced by the function analyzer are integrated to the prompts of both the writer agents (to guide fuzz driver generation) and analyzer agents (to prevent invalid modification suggestions).

\myparagraph{Integrating Crash Validation Agent} 
Similarly, we integrate the Crash Validation Agent to the end of the \OFG's pipeline, configuring it to execute after crashes, classified by the Crash Analyzer as ``Program Errors'', occurs. 
If crashes are validated as false positive crashes, we integrate the fix recommendation produced by the Crash Validation agent to the prompt of the the Enhancer agent in the next cycle, so as to refine the fuzz driver and prevent the occurrence of similar crashes.

\subsubsection{Improving \OFG's architecture}
\label{subsubsec:ofg-improvements}
We made two changes to \OFG's architecture to support the additional agents:

\myparagraph{Inter-agent Communication via the Shared Repository Pattern}
\OFG agents originally communicated using the pipe-and-filter pattern (\cref{subsec:context-oss-fuzz-gen}). 
However, sharing the function analyzer’s results with multiple agents across different pipeline stages was inefficient: each agent propagated the generated constraints (from its predecessor's result object) on to downstream agents for access.


To address this, we extended \OFG to support the Shared Repository Pattern~\cite{lalanda_shared_1998}.
Here, agents operating on the same function write results to a central repository accessible by others. 
Since agents execute in isolated cloud containers, the repository resides in the orchestrator’s working directory.
Each agent copies it when provisioned. 
After execution, any new or modified files are synchronized with the central repository, so that future agents can access shared data without intermediate pipeline transfers.
    
\myparagraph{Capture-and-Replay for Agent Debugging}
In multi-agent systems, downstream agents may depend on upstream outputs, making it hard to evaluate one agent without executing all preceding stages. 
This was especially challenging for the Crash Validation agent, which uses outputs from the execution stage and the Crash Analyzer agent. 
Prior work on designing multi-agent systems~\cite{li_survey_2024, pan_agentcoord_2025, gao_agentscope_2024, he_llm-based_2025} rarely cover design techniques that enable independent agent validation, so we describe our approach here.

We implemented a \textit{capture-and-replay} approach~\cite{jha_capture_2013} that enables independent execution and debugging of agents. 
In \OFG, agent inputs are embedded in prompts with clearly demarcated XML tags, which are logged. 
Our framework extracts these components to recreate the context preceding an agent’s execution, allowing rapid debugging, prompt iteration, and repeated evaluation of agents like the Crash Validation agent (\cref{subsubsec:method-rq2}). 
Similar capture-and-replay techniques have been applied to test other software infrastructure~\cite{liu_capture-replay_2014, leotta_capture-replay_2013, steven_jrapture_2000, deljouyi_understandable_2024}.

\section{Evaluation}
\label{sec:evaluation}
The two strategies proposed in this paper both address false positive crashes in \OFG, but operate in different phases of \OFG's pipeline, making them orthogonal approaches. 
We therefore evaluate each strategy separately within \OFG and assess the cost they introduce to \OFG.

\vspace{0.1cm}
\begin{itemize} [label={}]
\item \textbf{RQ1:} How effective is \textit{constraint-based fuzz driver generation} in reducing crashes in \OFG?
\item \textbf{RQ2:} How effective is \textit{context-based crash validation} in identifying false positives in \OFG?
\item \textbf{RQ3:} What is the additional cost introduced by LLM-based agents to \OFG?
\end{itemize}


\subsection{Experimental Setup}


\myparagraph{Datasets}  
\OFG provides a set of 1555 benchmark functions, drawn from 336 (out of the 1311) \OF projects, that we use to continuously evaluate \OFG.
Each benchmark function represents one function in the parent \OF project, for which we develop a fuzz driver.
We use the full set of benchmark functions for this evaluation.
TO avoid resource exhaustion using experiments, we divide the full set into three subsets (yielding 510, 510 and 535 benchmark functions), and separately execute \OFG on each subset.

\myparagraph{Compute Resources}  
All experiments were executed on compute clusters on Google Cloud and were conducted between July and September 2025. 
The LLM-based agents were powered by the Gemini 2.5 Pro, Google's state of the art reasoning model at the time.  

\myparagraph{Baseline}
To evaluate the impact of the introduced strategies, we used vanilla \OFG as a baseline.


\subsection{Methodology}

\subsubsection{Executing \OFG Experiments for Evaluation}

We evaluated two \OFG configurations: one where the generated function constraints were provided to \OFG agents, and one where they were ignored. 
Both configurations were executed on the three benchmark subsets. 
\OFG performed 10 trials for each benchmark function, with each trial generating and iterating on a fuzz driver for up to five cycles. 
Each driver was executed for five minutes in the execution stage.

\subsubsection{RQ1: Effectiveness of constraint-based fuzz driver generation}

We examined how constraint-based generation influences \OFG outcomes.  
We first compare total number of crashes and crashes reported as false positives across the two configurations and on the three benchmark sets. 
Additionally, to ensure difference in number of crashes is not caused by weaker code exploration, we also compare the average coverage achieved by fuzz drivers in each configuration.  

Next, we evaluate the impact of function constraints on fuzz driver quality by measuring how well generated drivers satisfy the constraints produced by the Function Analyzer. 
We evaluate this using the two \OFG{} configurations and the first benchmark set comprising 510 benchmark functions.
For each function in the benchmark set, we selected the first two trials, and excluded cases either constraint or fuzz driver generation failed. 
This yielded 957 test cases for the first configuration and 963 for the second. 
Using the Gemini 2.5 Pro model, we assessed constraint satisfaction and validated reliability by manually reviewing a random sample of five functions. 
We then report the proportion of fuzz drivers that fully satisfy all derived constraints.

Finally, we conduct a preliminary investigation into why constraint-based fuzz driver generation strategy was insufficient to fully mitigate crashes, and to provide insights to the limitations for this strategy.
We randomly sampled 20 crashes, 10 labeled as false positives and 10 as true positives, and analyzed why they weren't mitigated by the provided constraints.
We classify and report these reasons, together with the number of crashes belonging to them.

\subsubsection{RQ2: Effectiveness of context-based crash validation}  
\label{subsubsec:method-rq2}

We evaluated the impact of the Crash Validation agent in identifying additional false positive crashes reported by \OFG.  

First, for each benchmark set, we measure the number of crashes initially classified as program errors by the Crash Analyzer that were later marked as false positives by the Validation agent. 
This proportion reflects the impact of the additional constext-based crash validation stage is reducing the final number of false positives reported to maintainers.

Next, we assess the reliability of the crash validation agent's analysis and conclusion.
We randomly sample, review and characterize 20 agent interactions and final analysis and evaluate the extent to which they followed the steps prescribed to the agent in~\cref{subsubsec:cv-agent-design}.
Additionally, we measure how consistent the agent's conclusion is, across up to three executions of the same prompt, to estimate the agent's correctness.
This is because prior work has shown correlation between LLM consistency and correctness.
We randomly sample 200 crashes that we had previously determined its validity using the crash validation agent, and using the debugging framework in \cref{subsubsec:ofg-improvements}, we ran three repeated experiments, measured the proportion whose results were consistent, and investigated reasons for inconsistencies.

Finally, we evaluate the impact of providing detailed instructions to the crash validation agent.
Alongside the original prompt, we create a second version that only specifies the crash validation task but no decomposition step guidance, and rerun the crash validation agent with each prompt on these sampled subset of 200 crashes used above.
We compare differences in the agent's conclusions, tool usage, and output token.

\subsubsection{RQ3: Cost overhead of FalseCrashReducer} 
\label{subsubsec:rq3-methods}

We assessed the cost overhead introduced by the two LLM-driven strategies to \OFG.
In this RQ, we focus on LLM API costs. 
However, since the \OFG evaluation was executed on the Google Cloud Platform, it also incurred infrastructure costs which are more difficult to retroactively measure.

To measure cost, we extract all agent inputs and outputs for all benchmark functions and all executed agents during the evaluation experiment run.
We use the tokenizer tool from OpenAI to compute the number of input and output tokens and use the Gemini API pricing to estimate the cost of each agent on each benchmark.
Finally, we calculate and report the average cost of the default \OFG agents on each benchmark, and the average additional cost per benchmark introduced by the function analyzer agent and the crash validation agent.

\subsection{Results}


\subsubsection{RQ1: Effectiveness of constraint-based fuzz driver generation}  
\label{subsubsec:rq1-results}


\begin{table}
    \centering
    \ifTABLEFONTHACK
    \small
    \fi
    \caption{Crashes and coverage achieved by two \OFG configurations (baseline/without function constraints, and new/with FC).
    Parentheses denote number of false positive crashes. ``\# Bm'' denote number of benchmark functions.
    }
    \begin{tabular}{cP{0.7cm}P{1.5cm}P{1.5cm}P{0.7cm}P{0.5cm}P{0.5cm}}
        \toprule
        Set & \# Bm & \multicolumn{2}{c}{Num. Crashes} & \% diff & \multicolumn{2}{c}{Coverage}  \\
        \cmidrule(lr){3-4} \cmidrule(lr){6-7}
         &        & w/o FC & with FC &              & w/o FC & with FC \\
        \midrule
        Set-1 & 510 & 1858 (1307) & 1810 (1249) & 2.6\% & 22.5\% & 22.3\%  \\
        Set-2 & 510 & 1577 (1082) & 1450 (1026) & 8.1\% & 19.5\% & 19.0\%  \\
        Set-3 & 535 & 1645 (1194) & 1575 (1115) & 4.3\% & 20.3\% & 20.1\%  \\
        \midrule
        Total & 1555 & 5080 (3583) & 4835 (3390) & 15.0\% & 62.3\% & 61.4\% \\
        \bottomrule
    \end{tabular}
    \label{tab:rq1-crashes}
\end{table}

\cref{tab:rq1-crashes} shows that incorporating function constraints consistently reduced the number of total and false positive crashes across all benchmark sets, with reductions of up to 8.1\%.  
At the same time, fuzz drivers with constraints achieved very similar coverage, confirming that the reductions were not due to weaker code exploration.  

\begin{table}
    \centering
    \ifTABLEFONTHACK
    \small
    \fi
    \caption{Satisfaction of function constraints (FC) by generated fuzz drivers. 
    Drivers generated with FC satisfy notably more constraints compared to those without.}
    \begin{tabular}{lcc}
    \toprule
        Metric & w/o FC & with FC \\
        \midrule
        \# Fuzz Drivers Analyzed & 908 & 900 \\
        Avg constraints per driver & 4.33 & 4.25 \\
        \% satisfying all constraints & 38.9\% & 63.1\% \\
        \% satisfying $\geq (n-1)$ constraints & 68.7\% & 88.2\% \\
        Overall constraint satisfaction & 73.2\% & 86.9\% \\
        \bottomrule
    \end{tabular}
    \label{tab:rq1-constraints}
\end{table}

Furthermore, we also evaluated impact on fuzz driver quality, assessing how fuzz drivers conform to the derived function requirements.
\cref{tab:rq1-constraints} shows that fuzz drivers generated with constraints were substantially more likely to satisfy the expectations of the target function: 63.1\% satisfied all derived constraints compared to only 38.9\% without constraints.  
This indicates that explicitly providing function constraints significantly improves fuzz driver quality, even though \OFG’s writer agents already have access to the source code.

\begin{table}
    \centering
    \ifTABLEFONTHACK
    \small
    \fi
    \caption{Reasons why crashes were not mitigated by the constraint-based fuzz driver generation strategy.}
    \begin{tabular}{lc}
    \toprule
        Metric & Value \\
        \midrule
        Number of crashes studied & 20 \\
        \midrule
        Crashes occurred in non-analyzed function & 12 \\
        Crashes occurred beyond scope of analyzed function & 4 \\
        Crashes caused by incomplete constraints & 3 \\
        Crashes caused by conflicting suggestions & 1 \\
        \bottomrule
    \end{tabular}
    \label{tab:rq1-fc-limitations}
\end{table}

Finally, we examined the limitations of the constraint-based fuzz driver generation strategy. 
\cref{tab:rq1-fc-limitations} summarizes the causes of 20 sampled crashes. 
Twelve originated in other functions in the fuzz driver beyond the target function, and four more occurred in functions up to two levels deeper in the call graph from the target function.
All of these are outside the scope of the Function Analyzer. 
Three were due to incomplete constraints.
For example, a double free caused by a callback freeing an input pointer and the constraints did not specify expectations of callbacks provided to the target function. 
The final crash resulted from a coverage improvement suggestion from the coverage analyzer agent that directly violated a constraint. 
These findings highlight how the strategy can be expanded to better prevent crashes.

\subsubsection{RQ2: Effectiveness of context-based crash validation}  
\label{subsubsec:rq2-results}

\begin{table}
    \centering
    \ifTABLEFONTHACK
    \small
    \fi
    \caption{Percentage of crashes classified as ``Program Errors'' that were filtered out using context-based crash validation.}
    \begin{tabular}{lcccc}
    \toprule
        Metric & Set 1 & Set 2 & Set 3 \\
        \midrule
        Crashes caused by ``program errors'' & 1092 & 840 & 853 \\
        Additional false positives identified & 626 & 548 & 522 \\
        \% non-feasible crashes filtered & 57.3\% & 65.2\% & 61.2\% \\
        \bottomrule
    \end{tabular}
    \label{tab:rq2-fp}
\end{table}

\begin{listing}
  \centering
  \caption{Analysis from crash validation agent, showing why a specific crash is infeasible from the program's entry point.}
  \label{listing:flac_fp_analysis}
\begin{minted}[
    fontsize=\footnotesize,
    linenos,
    xleftmargin=0.5cm,
    style=colorful,
    breaklines,
    breaksymbolleft=,
    escapeinside=||
]{xml}

<feasible> False </feasible>
<analysis>
The crash is a heap-buffer-overflow read in `flexbuffers::GetRoot` caused by reading from a buffer with a size less than 2. The function reads the last two bytes of the buffer to determine the `byte_width` and `packed_type` without first checking if the buffer is large enough to support these reads. The fuzzer triggers this by calling `flexbuffers::GetRoot` with a 1-byte buffer.

The crash is not feasible... When `flatc` processes a FlexBuffer file, it invokes `flexbuffers::VerifyBuffer` before calling `flexbuffers::GetRoot`. 
|\textbf{This verification step ensures the buffer is at least 3 bytes long, which is a stricter check than required to prevent the crash (a 2-byte minimum).}| 
...
</analysis>
<source_code_evidence>...</source_code_evidence>
<recommendations>
...
</recommendations>
\end{minted}
\end{listing}

\cref{tab:rq2-fp} shows that the Crash Validation agent filtered 57–65\% of crashes initially flagged as program errors, highlighting the value of context-based crash analysis in reducing false positives.  
For example, \cref{listing:flac_fp_analysis} illustrates a case where the agent correctly deemed a crash infeasible: although triggered by a bug in the target function, the program's entry-point input validation will prevent this crash during normal execution.

\begin{table}
    \centering
    \ifTABLEFONTHACK
    \small
    \fi
    \caption{Reasons for unreliable crash validation analysis.}
    \begin{tabular}{lc}
    \toprule
        Metric & Count \\
        \midrule
        Crashes reviewed & 20  \\
        Incorrect localization of crash & 5 (25\%)\\
        Incomplete consideration of call sites & 2 (10\%)  \\
        Incorrect identification of root cause & 1 (5\%)  \\
        Incorrect identification of crash conditions & 1 (5\%)  \\
        Incorrect constraint and feasibility analysis & 1 (5\%) \\
        \midrule
        Analysis with reliable conclusions & 10 (50\%)  \\
    \bottomrule
    \end{tabular}
    \label{tab:rq2-verifiability}
\end{table}

\myparagraph{Evaluating reliability of agent's analysis}
Next, we manually reviewed 20 crash analyses (\cref{tab:rq2-verifiability}) to determine if they followed the prescribed validation steps (\cref{subsubsec:cv-agent-design}) and produced reliable results. 
Half were reliable, as they followed the steps and provided source code evidence to back all claims. 
In six cases, the agent mislocalized the crash or misidentified the root cause. 
In three others, it failed to fully analyze call sites or constraints; and in one case it misjudged the crash conditions. 
Of the reliable cases, four corresponded to real bugs and we are in the process of reporting them to the project maintainers.  

On further result inspection, errors from mislocalized crashes and misidentified root causes mostly stemmed from imprecise root cause analysis from the crash analyzer agent, which occurred because the crash analyzer either struggled to identify the actual root cause during its analysis or did not fully communicate the program flow that led to the crash, leading to assumptions in the validation agent. 

Overall, these results show that while the crash validation agent provides reliable conclusions half the time, its performance can be improved by more accurate analysis from upstream agents and providing it with tools that enable systematic validation of function call sites and constraints.

\begin{table}
    \centering
    \ifTABLEFONTHACK
    \small
    \fi
    \caption{Consistency of the Crash Validation agent's conclusions across 3 repeated runs. Inconsistencies are primarily due to conflicting usage assumptions and reachability of the crashing function from program entry points.}
    \begin{tabular}{lc}
    \toprule
        Metric & Value \\
        \midrule
        Crashes evaluated & 200 \\
        Consistent conclusions & 137 (68\%) \\
        Consistent false-positive calls & 130 (65\%) \\
        Consistent true-positive calls & 6 (3\%) \\
        \midrule
        Sampled inconsistency analysis investigated & 10 \\
        Inconsistencies in real-world usage assumptions & 5 \\
        Inconsistencies in crash function reachability & 4 \\
        Inconsistencies in identified root causes & 1 \\
        \bottomrule
    \end{tabular}
    \label{tab:rq2-consistency}
\end{table}

\myparagraph{Evaluating consistency of agent's conclusions}
We also evaluated the consistency of the validator's conclusions.
Across three repeated runs, 68\% of conclusions were consistent (\cref{tab:rq2-consistency}), with most consistent outcomes corresponding to false positive classifications.  
Following prior work showing correlation between LLM consistency and accuracy~\cite{valentin_estimating_2025, jiang_representation_2025}, this result provides a measure of the accuracy of the crash validation agent and show it can more accurately distingush false positives from true crashes.  

To further understand the reasons behind inconsistencies, we reviewed 10 sampled cases. 
Five arose from conflicting assumptions about real-world usage, four from differing reachability judgments, and one from root-cause disagreement. 
For instance, one analysis marked a crash feasible because a user-controlled variable could be zero, while another marked it infeasible because it assumed users would never set it to zero. 
Similarly, another crash was marked feasible by one analysis because it can be triggered by three public functions while a second analysis considered it infeasible because the public functions were correctly used within the project.
Detailed examples are provided in the supplemental materials.  
We observed that, even when analyses differed, they still provided evidence about the crash's feasibility to inform human debugging.

\myparagraph{Evaluating impact of problem decomposition steps in prompt}
Finally, we compared two prompt designs: one with explicit decomposition steps (\cref{subsubsec:cv-agent-design}) and a simple one without. 
Results diverged in 60\% of cases.
The detailed prompt yielded 16.3\% longer outputs on average (842 vs.\ 724 tokens) but produced a similar number of tool calls (6.9 vs.\ 6.2), suggesting greater verbosity without added efficiency.

\subsubsection{RQ3: Cost overhead of LLM-based agents}  
\label{subsubsec:rq3-results}

\begin{table}
    \centering
    \caption{Average per-driver cost of \OFG agents (USD), \ie applying these agents to generate and refine a driver for a single function within an \OF project. New agents adds roughly 9.31\% to the cost of existing agents. On 10 concurrent trials per function and 1555 functions from 336 projects, executing \OFG with the introduced agents cost on average, \$43.50 per evaluated project and \$14,616 for all 336 projects.}
    \begin{tabular}{lcccc}
    \toprule
        Agent & Input & Tools & Output & Total \\ 
        \midrule
        Function Analyzer & \$0.004 & \$0.016 & \$0.004 & \$0.024 \\
        Context Analyzer & \$0.022 & \$0.023 & \$0.012 & \$0.056 \\
        Existing agents & \$0.055 & \$0.685 & \$0.119 & \$0.859\\
        \midrule
        Cost (per driver) & \$0.081 & \$0.723 & \$0.135 & \$0.939 \\
        \midrule
        Cost (per project) & \$3.75 & \$33.49 & \$6.26 & \$43.50 \\
        \bottomrule
    \end{tabular}
    \label{tab:rq3-cost}
\end{table}


\cref{tab:rq3-cost} summarizes the financial overhead introduced by the Function Analyzer and Crash Validation agents. 
The Function Analyzer costs only \$0.024 per fuzz driver, substantially cheaper than other agents, while still reducing false positive crashes by 2–8\%.  
In contrast, the Crash Validation agent introduces more cost per driver, but has higher impact, eliminating over 50\% of false positive crashes reported as \textit{``Program Errors''}.  
Together, these agents increase \OFG API cost by about 9.31\% but contributes significantly to reducing false positive crashes and improving fuzz driver quality.  

We also observe that 68\% of \OFG’s total cost arises from tool interactions.  
This is largely due to the flexible code search tools provided to agents: retrieving source code for a single function may require multiple commands (\eg using \texttt{grep} to locate function definitions, then using \texttt{cat} to retrieve the entire file content).  
Providing more specialized tools could reduce the number of such interactions and lower costs further.  

\subsection{Threats to Validity}


We identify the various limitations of our work:

\ul{Construct Validity:}
As our work in built within the \OF context, our evaluations only reported fuzzing behaviors that led to crashes.
Hence, a low-quality fuzz driver can cause non-crash defects, which will not be captured by our evaluations. 
However, the fuzzing literature generally leaves other expected behaviors to the software owners to specify via asserts.

In RQ3 (\cref{subsubsec:rq3-methods}), we use API cost to approximate the cost of an agent-based approach.
This is imprecise, but note that the generated fuzz drivers are then run continuously, dwarfing generation costs.


\ul{Internal Validity:} 
The primary threat here is in the small sample sizes used in our detailed analysis of our agents' behaviors.
The detailed analyses reported here are consistent with our experience of agent behaviors during development and evaluation.

\ul{External Validity:}
We conducted our experiments using the Gemini 2.5 Pro model and the results may not generalize to other AI models.
Similarly, our evaluations were conducted using benchmark functions from \OF projects.
We did not characterize the evaluated functions and do not make claims of generalizability of functions that are more complex than the ones we evaluated or in projects such as embedded firmware or software applications which differs from the ``IT infrastructure'' class of projects on \OF.

\section{Lessons Learned and Open Problems}

\subsection{On Constraint-based Driver Generation}

Function constraints helped fuzz drivers satisfy target function expectations and modestly reduced false positives (\cref{subsubsec:rq1-results}), but crashes remain frequent, averaging four per function.  

Many stem from functions outside the analyzer’s scope, where no constraints can be generated. 
Extending \OFG with real-time constraint retrieval during driver generation, inspired by advances in context-aware code generation~\cite{zhang_codeagent_2024, wang_coderag-bench_2025, nashid_contextual_2024, jain_mitigating_2025, chen_when_2025}, could reduce these crashes but introduce time and cost overhead.

These false positives arise when constructing valid inputs and states for intermediate functions. 
The reader may therefore suggest focusing on public entry points, where drivers are easier and more reliable, while using directed fuzzing~\cite{bohme_directed_2017, fang_ddgf_2024, luo_selectfuzz_2023, canakci_targetfuzz_2022}, program state restriction~\cite{srivastava_one_2022}, and smarter seeds~\cite{shi_harnessing_2024, wang_skyfire_2017, lyu_smartseed_2019} to reach deeper functions. 
While this approach is complementary to our proposals and can cover deeper functions with poor coverage, it involves executing all intermediate functions between the entry point and the target function and may be inefficient when the target function is far from the entry point.  
Empirical measurements would be of interest, to compare
  the overhead cost of mitigating false postivies during bottom-up fuzzing
  with
  the reduction in efficiency introduced by directed top-down fuzzing.


\subsection{On Context-based Crash Validation}

Context-based validation removed over half of false positives but remained limited by inaccurate root-cause analysis and unsystematic callsite analysis (\cref{subsubsec:rq2-results}). 
This raises two key questions.  

First, should agents in multi-agent systems incorporate confidence or trust in upstream results? 
If prior analysis is uncertain, downstream agents could discount or discard it. 
While recent work explores LLM confidence estimation~\cite{becker_cycles_2024, xiong_can_2024, farr_llm_2024, tripathi_confidence_2025}, its relevance to program analysis is unexplored.  

Second, can lightweight program analysis tools improve validation? Adding call-graph queries~\cite{hall_efficient_1992} or program slicing~\cite{xu_brief_2005} may mitigate unsystematic reasoning, but static analysis is often imprecise~\cite{walker_limitations_2020}. 
Better designs may balance innovative agent architectures with minimal tooling to produce more reliable results.  

Finally, the consistencies observed in repeated agent analysis highlight the complexity of crash validation and the need for empirical research to guide distinguishing true from false crashes.
In addition, they also demonstrate the need for stronger crash validation methods, such as building the complete buggy program and generating entry-level inputs to trigger and validate the crash.
This line of work can build on prior exploit generation works~\cite{xu_automatic_2018, you_semfuzz_2017, peng_pwngpt_2025}.

\subsection{Cost of an \OFG Approach}

\OFG costs stem mainly from API usage and compute costs.
However, compute costs are negligible when compared to the cost of continuous fuzzing.  
As shown in \cref{subsubsec:rq3-results}, generating a fuzz driver costs under \$1 in API usage, averaging about \$43.50 per project across the evaluated set.  
Generated drivers achieve 8.4\% coverage on average, with some projects reaching 98\%.  

These costs are minimal compared to manual development.  
The \OF program pays up to \$15,000 for integrations reaching 50\% coverage~\cite{noauthor_oss-fuzz_nodate}, estimatedly about \$2,400 for the 8\% average coverage of \OFG.  
Despite such incentives, contributions remain low due to required expertise.  
Were Google employees to do the development work, consider that a junior Google engineer is paid $\sim$\$70/hour in salary~\cite{noauthor_google_levelsfyi_nodate}.
\OFG generates 10 drivers and 8\% coverage for roughly \$35, or less than an hour of developer time. 

However, there are still opportunities to reduce API costs.
For example, tool usage accounts for 77\% of these costs, often because retrieving program symbols or function callsites involves multiple tool queries (\eg using \texttt{grep} to find symbol location and \texttt{cat} to extract code snippets).
Hence, streamlined tools that provide easy access to program symbols and call graph will reduce number of tool invocations and API cost.


\subsection{Implications for Research and Practice}

FalseCrashReducer advances the practicality of \OFG and fuzz driver generation.  
By automating driver creation and crash validation, it accelerates fuzzing and enables new directions:

\myparagraph{Debugging assistance}
The Function Analyzer and Crash Validation agents, though integrated with \OFG, are loosely coupled and can be adapted for other fuzz driver generation or automated testing systems. 
By sharing these agents, their evaluation results, and discussions of their strengths and limitations, we provide guidance for future research and offer raw data as measurement baselines. 
We also observed that even imperfect agent analyses offer valuable insights for debugging and crash triaging. 
Future work could empirically assess how these imperfect analyses affect the effort and time required to debug and fix fuzzer crashes.

\myparagraph{CI/CD Integration}  
With drivers generated for every function and bugs uncovered within minutes, bottom-up fuzzing could be integrated into CI/CD pipelines.  
Unlike existing work that focus on top-down fuzzing~\cite{klooster_continuous_2023}, however, scaling bottom-up fuzzing may overwhelm pipelines and compute.  
Research is needed on novel deployment strategies: which functions to fuzz, how often and how long, and how engineers should interact with results.

\myparagraph{AI-Assisted Bug Fixing}  
Faster fuzzing will surface more crashes than maintainers can fix, as seen in \OF~\cite{serebryany_oss-fuzz_2017} and Syzkaller~\cite{noauthor_googlesyzkaller_2025}.  
Automated fixing with LLMs is promising~\cite{zhang_fixing_2024} but underexplored.  
Future work should classify which crashes are tool-fixable versus human-required, and empirically develop metrics to estimate reliability of AI-generated patches from crash features.
This will enable incremental adoption of AI-assisted bug fixing and lead to stronger support for fuzz driver generation.

\myparagraph{Bounty Program Impact}  
As automated driver generation spreads, bounty programs~\cite{noauthor_oss-fuzz_nodate} may see more reports, including an increase in false positives, just as project maintainers already report frustration with low-quality AI-generated bug reports~\cite{aaron_death_2025, noauthor_open_nodate}.  
Hence, bounty policies may need stricter validation, stronger proof-of-concepts, or new criteria to balance discovery with maintainability.

\myparagraph{Advancing Techniques with Stronger Guarantees}  
We can apply improvements in fuzzing automations to automate other verification methods like bounded model checking~\cite{clarke_bounded_2001}.  
Formal methods provide stronger guarantees but require expertise.  
While recent work show progress on reducing memory safety verification cost~\cite{amusuo_unit_2025-1, amusuo_unit_2025}, novel AI-assisted automation could further lower barriers to formal methods.  
With potential improvements in usability of formal methods techniques, future work should also explore how fuzzing and formal methods complement one another and improve the scalability of underlying formal methods tools like SMT solvers.

\section{Conclusion}
False positive crashes remain a major challenge in fuzz driver generation and bottom-up fuzzing.
We proposed two novel agent-driven strategies, implemented and integrated into \OFG, to mitigate this problem.
Our evaluation shows a modest improvement in reducing false positive crashes and substantial improvement in filtering crashes after the event.
Further analysis quantifies their reliability, consistency, and cost, providing evidences of their strengths and weaknesses.
Overall, these strategies mark a significant step toward practical, industry-scale fuzz driver generation, directly benefiting critical open-source projects on \OF.

\section*{Open Science}

All systems described in this paper are open-source:
  \url{https://github.com/google/oss-fuzz-gen}.
For evaluation scripts and raw data, see: \url{https://github.com/PurdueDualityLab/ICSE-SEIP26-FalseCrashReducer}.

\begin{acks}
    We thank Dustin Ingram for his contributions.
    Davis acknowledges support from Rolls Royce and Qualcomm.
\end{acks}


\bibliographystyle{ACM-Reference-Format}
\bibliography{main}

\end{document}